\documentclass[twocolumn,superscriptaddress,secnumarabic,amssymb, nobibnotes, aps, bm,prl,floatfix,longbibliography]{revtex4-2}
\usepackage[english]{babel}
\usepackage[latin9]{inputenc}
\usepackage[usenames,dvipsnames]{xcolor}
\usepackage{mathtools}
\usepackage{amsmath}
\definecolor{goodgreen}{rgb}{0.1,0.5,0}
\definecolor{goodred}{rgb}{0.7,0,0}
\usepackage[colorlinks,urlcolor=goodgreen,citecolor=blue,linkcolor=goodred]{hyperref}
\usepackage{cleveref}

\makeatletter

\usepackage{soul}

\usepackage{braket}

\usepackage[normalem]{ulem}

\newcommand{\beq}{\begin{equation}}
\newcommand{\eeq}{\end{equation}}
\newcommand{\bea}{\begin{eqnarray}}
\newcommand\bal{\begin{aligned}}
\newcommand\eal{\end{aligned}}
\newcommand{\eea}{\end{eqnarray}}

\makeatother
\begin{document}
\title{Topological Phases for Extended Objects:\\ Semiclassical Phase-Space Approach with Tensorial Coordinates}

\author{Giandomenico Palumbo}
\email{giandomenico.palumbo@gmail.com}
\affiliation{School of Theoretical Physics, Dublin Institute for Advanced Studies, 10 Burlington Road, Dublin 4, Ireland}

\begin{abstract}
\noindent  It has long been established that certain higher-dimensional topological phases of matter support extended objects like quasi-strings and quasi-membranes in their bulk states. In this study, we investigate the physics of these topological systems using a phase-space approach in the semiclassical regime, incorporating tensorial coordinates related to the extended objects. Specifically, we explore the semiclassical currents associated to open quasi-strings in topological phases in three spatial dimensions, considering the presence of both coordinate-space Kalb-Ramond fields and momentum-space tensor Berry connections. Our results show that the currents associated to the endpoints of open quasi-strings on the system's boundary exhibit a quantized Hall response. This formalism accurately reproduces the boundary response as predicted by an Abelian BF theory proving a novel way to study extended objects in topological matter.
\end{abstract}
\date{\today}
\maketitle

\noindent \section*{\bf{Introduction}}
\noindent Topological matter represents one of the most active research fields in modern physics, mainly due to its interdisciplinarity \cite{Wen}. It serves as an ideal playground where ideas and techniques from various areas, ranging from high-energy physics and quantum information to solid-state physics and ultracold atoms, converge to create novel theoretical paradigms and experimental discoveries \cite{Zhang,Haldane}.
Although most of the topological phases of matter discussed in the literature involve free or interacting microscopic systems composed of point-like quasi-particles (such as fermions, bosons, and anyons), they also encompass more exotic physical systems. These systems, primarily in three or higher spatial dimensions, support extended objects such as strings and membranes. This phenomenon is not limited to string theory \cite{Tong}, but it also occurs in certain condensed matter systems where emergent extended objects are induced by interactions and geometric configurations of more fundamental point-like objects.
For example, emergent superconducting/superfluid vortex strings \cite{Rasetti-Regge,Lund-Regge,Davis,Franz}, skyrmion strings \cite{Yokouchi,Kravchuk} and membranes \cite{Tizzano,Palumbo2022} are well-known cases of extended objects that can, in principle, be described in the low-energy limit through the language of quantum field theory. Their potential topological features are effectively encoded in suitable effective field theories.
More specifically, within the framework of gauge theory, topological actions coupled to extended-object currents can be naturally constructed by introducing higher-form (Kalb-Ramond) gauge connections \cite{Kalb} such as in the case of BF theories in three spatial dimensions \cite{Horowitz,Blau,Guadagnini,Zee,Orland,Semenoff,Amoretti}. BF actions, possibly augmented by Maxwell-like kinematic terms for the one- and two-form fields, have been shown to describe several topological phases \cite{Thouless,Sodano,Marzuoli, Fradkin,Senthil,Cirio,Simon, Ye,Cirio2,You,Tiwari,Tiwari2,Putrov,Randellini,Cheng,Palumbo2020,Grushin,Pace}, similar to higher-order Chern-Simons theories that can describe higher-dimensional quantum Hall states for extended objects \cite{Tizzano,Palumbo2022}. 
Although space-time-dependent effective actions can describe both the topological and dynamical aspects of topological matter, they do not usually capture all of their fundamental aspects, such as emergent vector Berry connections in momentum space, which are crucial for characterizing many topological phases \cite{Niu}. Therefore, phase-space topological field theories are more convenient and efficient theoretical tools, as emphasized in some previous works \cite{Bulmash,Hayata}. Moreover, we briefly summarize additional reasons to consider a phase-space framework in topological matter.
Firstly, in several microscopic topological systems, the corresponding momentum-space Berry connections not only allow us to derive their topological invariants but also significantly influence the dynamics of quasi-particles \cite{Niu,Niu2}. At a semiclassical level, this is evident through a phase-space approach, where electrons experience the influence of external electric and magnetic fields, as well as static or dynamical Berry connections that modify the symplectic structure of the phase-space in a non-trivial way \cite{Niu3,Duval,Northe}. Secondly, at a fully quantum level, the phase-space approach allows us to employ deformation quantization (Wigner-Weyl quantization), replacing the Poisson brackets of the classical variables with Moyal brackets involving the quantum observables \cite{Zachos}. One of the main benefits of deformation quantization in condensed matter physics is its correct description of the bulk states of fractional quantum Hall phases, which are incompressible and invariant under quantum area-preserving diffeomorphisms \cite{GMP,Karabali,Cappelli1993,Palumbo2023}. This quantum-geometric feature, encoded in Moyal brackets and corresponding emergent non-commutative geometry, predicts not only the existence of a spin-2 (graviton-like) collective massive mode \cite{Yang}, but also bosonic \cite{Gromov,Cappelli2016,Palumbo2024} and fermionic \cite{Salgado-Rebolledo,Nguyen} higher-spin massive modes  in fractional quantum Hall states.
Thus, it is highly desirable to employ a phase-space approach for extended objects by generalizing the previous results obtained for quasi-particles.\\
In this work, we start this exploration in a semiclassical limit, in which the extended objects are coupled to both external higher-form gauge fields and tensor Berry connections. The latter can be seen as momentum-space higher-form gauge fields and have been originally introduced by the author to study novel topological phases in three and four spatial dimensions \cite{Palumbo2018,Palumbo2019,Zhu2020,Palumbo21,Zhu2021,Chen}. For simplicity, we will focus on topological phases in three spatial dimensions that support quasi-strings (i.e., one-dimensional extended objects), but all the main results derived below can be naturally generalized to higher-dimensional extended objects.
We will then derive the corresponding semiclassical Hamilton equations through the phase-space formalism by considering antisymmetric tensorial coordinates for the quasi-strings, previously introduced in mathematics \cite{Givens} and independently in different research fields in high-energy physics \cite{Hofman,Carlson,Amorin,Bandos,Fed,Luk}. This peculiar choice of coordinates is crucial for deriving a complementary theory to the Hamilton-Jacobi formalism for strings proposed by Nambu \cite{Nambu,Eguchi,Rinke,Hosotani}, which instead involves tensorial momenta. We will briefly introduce the Nambu's approach before moving our complementary way in order to deal with open quasi-strings in phase space, by also including their endpoints.
As one of the main results of our investigation, we will show that the endpoints of quasi-strings, which can be seen as point-like quasi-particles living on the boundary of three-dimensional topological systems, give rise to a quantized Hall current.

\section*{Chern-Simons theory and semiclassical approach for quasi-particles in two spatial dimensions}

\noindent In this section, before addressing the main topic of this paper concerning extended objects, we provide a brief overview of the key features of the phase-space approach for quasi-particles that support a quantized Hall current \cite{Niu} and its agreement with a topological quantum field theory description. Notably, the integer quantum Hall effect can be described by an Abelian Chern-Simons theory, given by
\begin{eqnarray}\label{CS}
	S_{CS}=\int d^{3}x\,\left( \frac{\nu}{4 \pi}\,\epsilon^{\mu\nu\lambda}A_{\mu}\partial_{\nu}A_{\lambda}-J^{\mu}A_{\mu}\right),
\end{eqnarray}
where $\mu,\nu,\lambda=t,x,y$ is the spacetime index, $A_{\mu}$ is the U(1) gauge connection, $J^{\mu}$ is the electric current of quasi-particles and $\nu$ is the the first Chern number that coincides with the level of the Chern-Simons theory. By varying this action with respect to the gauge field, we can derive the covariant Hall current
\begin{eqnarray}
	J^{\mu}=\frac{\nu}{2 \pi}\epsilon^{\mu\nu\lambda}\partial_{\nu}A_{\lambda},
\end{eqnarray}
that can be rewritten by spitting the time-like and space-like components as follows
\begin{eqnarray}
	J^{i}=\frac{\nu}{2 \pi}\epsilon^{ij}E_{j}, \hspace{0.3cm} \rho=\frac{\nu}{2 \pi}B_{z},
\end{eqnarray}
where $i,j=\{x,y\}$, $B_{z}$ is the magnetic field that points in the z direction and we can recognize the transverse conductivity tensor, given by
\begin{eqnarray}
	\sigma^{ij}=\frac{\nu}{2 \pi}\epsilon^{ij}.
\end{eqnarray}
We now show that the conductivity tensor can be derived from a phase-space approach in the semiclassical regime \cite{Niu,Duval}. It has already been demonstrated that this method can identify the main topological features of various quantum systems in solid-state physics and cold atoms \cite{Cooper}, ranging from topological Bloch oscillations  \cite{Liberto} to charge pumping \cite{Price}. Here, we briefly summarize its main features for quasi-particles in two spatial dimensions. Point-like particles in real space are described by the coordinates $r_{i}=r_{i}(t)$, where $i$ is the space index and $t$ is time, the only parameter we have for the world-line trajectories of quasi-particles.
Their tangent vector, namely the velocity is 
\begin{eqnarray}
	v_{i}(t)=\dot{r}_{i},
\end{eqnarray}
while the corresponding conjugate momenta are given by 
\begin{eqnarray}
	p_{i}=\frac{\partial L}{\partial \dot{r}_{i}},
\end{eqnarray}
where $L$ is the Lagrangian of the system and in the Hamiltonian formalism, $r_{i}$ and $p_{i}$ are considered independent variables in the phase space.
Similarly, point-like quasi-particles in momentum space can be described by $p_{i}=p_{i}(t)$. In this dual picture, the corresponding conjugate momenta are given by $r_{i}=\partial L/\partial \dot{p}_{i}$, where now $L$ is defined in the momentum space.
We can then introduce the effective semiclassical action in phase space \cite{Duval}, which is based on the Faddeev-Jackiw formalism \cite{Faddeev-Jackiw}
\begin{eqnarray}\label{semi-classical-points}
	S_A=\int dt\, \left[(p_{i}-A_{i}(r))\dot{r}^{i}-(r_{i}-\tilde{A}_{i}(p))\dot{p}^{i}-H\right],
\end{eqnarray}
where $A_{i}(r)$ is an external electromagnetic potential, $\tilde{A}_{i}(p)$ is the momentum-space Abelian vector Berry connection and H($r_{i}$, $p_{i}$) is the microscopic Hamiltonian. Importantly, both the coordinate-space gauge fields and Berry connections are minimally coupled to the conjugate variables through the following substitutions
\begin{eqnarray}
	p_{i}\rightarrow p_{i}-A_{i}, \hspace{1.0cm} r_{i}\rightarrow r_{i}-\tilde{A}_{i}.
\end{eqnarray}
The equations of motion are then given by
\begin{eqnarray} \label{QHE}
	\dot{r}^{i}=\frac{\partial H}{\partial p_{i}}-\Omega^{ij}\dot{p}_{j}, \nonumber \\
	\dot{p}^{i}=-\frac{\partial H}{\partial r_{i}}-F^{ij}\dot{r}_{j},
\end{eqnarray}
where $F^{ij}=\partial^{i}A^{j}-\partial^{j}A^{i}$ is the magnetic component of the Faraday tensor (i.e. the magnetic field) while $\Omega^{ij}=\partial^{i}\tilde{A}^{j}-\partial^{j}\tilde{A}^{i}$ represents the time-independent Abelian Berry curvature.
We can now select an Hamiltonian H that reads
\begin{eqnarray}
	H(r^i, p^{i})=\mathcal{E}(p^{i})+V(r^i),
\end{eqnarray}
where $i=\{x,y\}$, $\mathcal{E}(p^{i})$ contains the usual kinematic term for non-interacting quasi-particles on a plane together with possible mass-like terms, while $V(r^i)$ is a potential that depends on the space coordinates, namely
\begin{eqnarray}
	V(r^i)=r^{i}E_{i},
\end{eqnarray}
where $E_{i}$ are the constant components of an external electric field. Importantly, here we also assume that H represents a microscopic Hamiltonian associated to an anomalous quantum Hall state (in a fermion system, this corresponds to consider, for instance, a continuum massive Dirac model that breaks time-reversal symmetry \cite{PalumboCigar,Para,Menezes}) such that the system supports a non-zero quantized first Chern number.
We can now combine the two expressions in Eq. (\ref{QHE}) and in the limit $F^{ij}\approx 0$, we get
\begin{eqnarray} \label{currentss}
	\dot{r}^{i}= \Omega^{ij} E_{j} +...,
\end{eqnarray}
where we have omitted to write the term $\partial H/ \partial p_i$ because we are mainly interested in the derivation of the topological response of the system. 
By bearing in mind the definition of vector current
\begin{eqnarray} \label{current2}
	J^{i}=\frac{1}{(2\pi)^2}\int d^{2}p\, \dot{r}^{i},
\end{eqnarray}
and by integrating both sides of Eq. (\ref{currentss}) on the two-dimensional compact momentum space, we finally obtain
\begin{eqnarray}
	J^{i}=\frac{\nu}{2 \pi}\,\epsilon^{ij}E_{j},
\end{eqnarray}
which coincides with the Hall current derived previously through the Chern-Simons theory, where
\begin{eqnarray}
	\nu=\frac{1}{2 \pi}\int d^{2}p\, \epsilon_{ij}\Omega^{ij}
\end{eqnarray}
is indeed the first Chern number. \\
\noindent In the next section, we will demonstrate that the first term in Eq. (\ref{semi-classical-points}) continues to play an important role in the semiclassical formalism for extended objects. Additionally, we will show that novel terms in a generalized effective action can be constructed in three spatial dimensions to establish a proper phase-space formalism for quasi-strings.

\section*{BF theory and semiclassical approach for quasi-strings in three spatial dimensions}

\noindent Here, we provide a novel semiclassical characterization of topological phases in three spatial dimensions that support quasi-strings in their bulk states. Similar to the previous section, we first derive topological currents from an effective action and then demonstrate how these topological responses can be derived directly within a phase-space formalism for extended objects.
We start considering an Abelian BF theory augmented by a $BB$ potential term that breaks time reversal symmetry \cite{Simon,Ye}. The corresponding action is then given by
 \begin{eqnarray}\label{BF}
	S_{BF}=
	\int d^{4}x\,\left( \frac{\kappa}{8 \pi}\,\epsilon^{\mu\nu\lambda\delta}\,\mathcal{B}_{\mu\nu}F_{\lambda\delta}-\frac{\kappa}{16 \pi} \epsilon^{\mu\nu\lambda\delta}\mathcal{B}_{\mu\nu}\mathcal{B}_{\lambda\delta} \right. \nonumber \\- \left. J^{\mu}A_{\mu}+\frac{1}{2}\,J^{\mu\nu}\mathcal{B}_{\mu\nu}\right), \,\,
\end{eqnarray}
where $\kappa$ is the quantized level of the action and coincides with a three-dimensional winding number (i.e. it is a $\mathbb{Z}$ invariant), $F_{\lambda\delta}=\partial_{\lambda}A_{\delta}-\partial_{\delta}A_{\lambda}$, with $\mathcal{B}_{\mu\nu}$ and $A_{\mu}$ two Abelian gauge fields and $J^{\mu\nu}=-J^{\nu\mu}$ is an anti-symmetric tensor current.
While $J^{\mu\nu}$ indicates that the two-form field couples to quasi-strings, $J^{\mu}$ refers to the vector current associated to their endpoints in the case of open quasi-strings.
This can be shown by integrating out $\mathcal{B}_{\mu\nu}$ in the action, yielding
\begin{eqnarray}
	S_{BF}=
	\int d^{4}x\,\left( \frac{\kappa}{16 \pi}\,\epsilon^{\mu\nu\lambda\delta}\,F_{\mu\nu}F_{\lambda\delta}+\frac{ \pi}{4\kappa}\,\epsilon^{\mu\nu\lambda\delta}\,J_{\mu\nu}J_{\lambda\delta} \right.  \nonumber \\- \left. J^{\mu}A_{\mu}+J^{\mu\nu}\partial_{\mu}A_{\nu} \right). \,\,
\end{eqnarray}
After integrating $A_{\mu}$, we find
\begin{eqnarray}
	\partial_{\mu}J^{\mu\nu}=J^{\nu},
\end{eqnarray}
which implies that the vector current associated to the propagation of quasi-particles actually refers to the endpoints of the quasi-strings. For closed quasi-strings, we instead find $\partial_{\mu}J^{\mu\nu}=0$.
Moreover, the first term in the action is a total derivative and induces the Abelian Chern-Simons term in Eq. (\ref{CS}) on the two-dimensional spatial boundary of the system, reproducing a boundary Hall-like response, similar to certain fermion topological systems in three space dimensions like topological insulators with gapped boundary \cite{Shen}. Thus, we assume in the rest of this section that the endpoints of the open quasi-strings live on the same planar boundary of the system and support a quantized Hall response.
This result can be derived in a complementary way by first varying the action in Eq. (\ref{BF}) with respect to $A_{\mu}$
such that we obtain the following bulk response
\begin{eqnarray}\label{stringQHE}
	J^{i}=\frac{\kappa}{2 \pi}\, \epsilon^{ijk}E_{jk}, \hspace{0.3cm} \rho=\frac{\kappa}{2 \pi}\, B_{w},
\end{eqnarray}
where $i,j,k=\{x,y,z\}$ and
\begin{eqnarray}\label{Etensor}
	E_{jk}=\mathcal{H}_{0jk}, \hspace{0.3cm} \mathcal{B}_w= \epsilon_{w ijk} \mathcal{H}^{ijk},
\end{eqnarray}
which are the tensorial electric field \cite{Henneaux} and the generalized magnetic field pointing in the fourth w-space direction associated to the curvature tensor $\mathcal{H}_{\mu\nu\lambda}$ of the Kalb-Ramond field $\mathcal{B}_{\mu\nu}$, i.e.
\begin{eqnarray}
\mathcal{H}_{\mu\nu\lambda}=\partial_{\mu}\mathcal{B}_{\nu\lambda}+\partial_{\nu}\mathcal{B}_{\lambda\mu}+\partial_{\lambda}\mathcal{B}_{\mu\nu}.
\end{eqnarray}
By defining the current $J^{i}$ on the $x-y$ flat boundary such that it no longer depends on the z-coordinate, we obtain
\begin{eqnarray}
	J^{i}=\frac{\kappa}{2 \pi}\, \epsilon^{ijz}E_{jz} \equiv \frac{\kappa}{2 \pi}\, \epsilon^{ij}E_{j},
\end{eqnarray}
with $E_{j}$ being the dimensionally reduced vector electric field, consistent with the quantized Hall-like response of the endpoints of the open quasi-strings.\\
We are now ready to derive a similar result by generalizing the semiclassical approach in phase space for extended objects. Our starting point is based on Nambu's proposal to describe bosonic strings in terms of Hamilton-Jacobi equations, where, unlike point-like quasi-particles, the conjugate momenta of extended objects are tensors \cite{Nambu,Eguchi,Rinke,Hosotani}. The coordinates for the worldsheets $r^{i}=r^{i}(t,s)$ depend on two independent parameters $t$ (time-like) and $s$ (space-like), characterizing the spacetime embedding of (open and closed) strings.
The tangent bivector for the worldsheet is given by
\begin{eqnarray}
	\mathring{r}^{ij}(t,s)=\partial_{t}r^{i}\partial_{s}r^{j}-\partial_{t}r^{j}\partial_{s}r^{i},
\end{eqnarray}
which allows us to define the conjugate tensorial momenta
\begin{eqnarray}
	P^{ij}=\frac{\partial L}{\partial\mathring{r}^{ij}},
\end{eqnarray}
where L is, for instance, the effective Lagrangian of the Nambu-Goto action \cite{Nambu}. In this way we have defined the independent phase-space variables $r^{i}$ and $P^{ij}$ of quasi-strings. Additionally, by considering a coordinate-space Kalb-Ramond field \cite{Kalb}, the effective action reads
\begin{eqnarray}\label{string-action}
	S_B=\int dt ds \left[(P_{ij}-\mathcal{B}_{ij}(r))\mathring{r}^{ij}-H(r^{i}, P^{ij})\right],
\end{eqnarray}
where $H(r^{i}, P^{ij})$ is an Hamiltonian for (closed) quasi-strings and we have included the minimal coupling \cite{Rey}
\begin{eqnarray}
	P_{ij}\rightarrow P_{ij}-\mathcal{B}_{ij}.
\end{eqnarray}
Note that $\mathcal{B}_{ij}$ term in $S_B$ in the symmetric gauge, i.e. $\mathcal{B}_{ij}=(1/3!)\epsilon_{ijk}r^k$ (such that its curvature tensor is constant), was originally introduced in Refs. \cite{Rasetti-Regge,Lund-Regge} although without the $P_{ij}$ term. 
By varying $S_B$ with respect to tensorial momenta, we get
\begin{eqnarray}
\mathring{r}^{ij}\equiv\{r^i, r^j\}_{st}=\frac{\partial H}{\partial P_{ij}},
\end{eqnarray}
where $\{, \}_{st}$ identifies the Poisson bracket of the worldsheet coordinates \cite{Chu}.
Within this framework, as a concrete example, we could consider the following effective Hamiltonian for quasi-strings
\begin{eqnarray}
	H(r^{i}, P^{ij})=\frac{1}{2}(P_{ij}-\mathcal{B}_{ij}(r))^{2},
\end{eqnarray}
with $i,j=\{x,y,z\}$, that formally generalizes to three space dimensions the well known planar Landau Hamiltonian for non-relativistic electrons coupled to an external electromagnetic field.
However, to derive the correct current $J^i$, previously obtained from the BF theory, we need to take a further theoretical step.
We propose a dual picture for the quasi-strings, starting from the standard momentum space. In this dual picture, the momentum-space coordinates for the worldsheet of the strings are defined by $p_{i}=p_{i}(t,s)$, and the corresponding tangent bi-vector is given by
\begin{eqnarray}
	\mathring{p}^{ij}(t,s)=\partial_{t}p^{i}\partial_{s}p^{j}-\partial_{t}p^{j}\partial_{s}p^{i},
\end{eqnarray}
which allows us to define the conjugate tensorial coordinates
\begin{eqnarray}
	X^{ij}=\frac{\partial L}{\partial\mathring{p}^{ij}},
\end{eqnarray}
with $X^{ij}=-X^{ji}$. Importantly, tensorial coordinates in real space have already introduced both in mathematics \cite{Givens} and independently in high-energy physics in the context of superstrings and noncommutative geometry \cite{Hofman,Carlson,Amorin,Bandos,Fed,Luk}.
In this dual picture, $p^{i}$ and $X^{ij}$ are the independent variables in the novel phase space, and the corresponding dual action reads
\begin{eqnarray}\label{string-action2}
	S_B^{dual}=\int dt ds \left[-(X_{ij}-\tilde{\mathcal{B}}_{ij}(p))\mathring{p}^{ij}-H(p^{i}, X^{ij})\right], \,
\end{eqnarray}
where a minimal coupling with respect to the tensorial coordinates
\begin{eqnarray}
	X_{ij} \rightarrow X_{ij}-\tilde{\mathcal{B}}_{ij}(p),
\end{eqnarray}
has been introduced, although now $\tilde{\mathcal{B}}_{ij}$ represents a momentum-space Abelian tensor Berry connection  \cite{Palumbo2019}, which is the natural generalization of vector Berry connections introduced in the previous section.
Because we aim to study the semiclassical topological features of open quasi-strings, we need to combine in our phase-space formalism both the phase-space variables of the quasi-strings and the conjugate variables of their endpoints.
For this reason, we propose the following semiclassical action
\begin{eqnarray}
	S_{AB}=
	\int dt ds \left[(p_{i}-A_{i}(r))\dot{r}^{i}-(X_{ij}-\tilde{\mathcal{B}}_{ij}(p))\mathring{p}^{ij} \right. \nonumber \\
\left.	-H(p^{i},X^{ij})\right],
\end{eqnarray}
with $F_{ij}(r)=0$, such that the corresponding generalized Hamilton equations with respect to $p^{i}$ and $X^{ij}$ are given by
\begin{eqnarray}\label{string-equations}
	\dot{r}^{i}=\frac{\partial H}{\partial p_{i}}-\tilde{\mathcal{H}}^{ijk}\mathring{p}_{jk}, \nonumber \\
	\mathring{p}^{ij}=-\frac{\partial H}{\partial X_{ij}}, \hspace{0.9cm}
\end{eqnarray}
where $\tilde{\mathcal{H}}^{ijk}$ is the curvature tensor of $\mathcal{\tilde{B}}_{ij}(p)$.
We can now consider an Hamiltonian H given by
\begin{eqnarray}
	H(p^{i},X^{ij})=\mathcal{E}(p^{i})+V(X^{ij}),
\end{eqnarray}
where $i,j=\{x,y,z\}$, $\mathcal{E}(p^{i})$ contains a kinematic term along with possible mass-like terms, and $V(X^{ij})$ is a potential that depends on the tensorial coordinates. This potential actually includes the appropriate external probing field, namely
\begin{eqnarray}\label{potential}
	V(X^{ij})=X^{ij}E_{ij},
\end{eqnarray}
where $E_{ij}$ is a constant tensorial electric field similar to that one previously introduced in Eq. (\ref{Etensor}).
Importantly, here we also assume that H breaks time-reversal symmetry and represents an effective Hamiltonian characterized by a topological phase in three spatial dimensions that supports a quantized three-dimensional winding number $\kappa$ (i.e. a first Dixmier-Douady invariant \cite{Palumbo2019}) given by
\begin{eqnarray}
	\kappa=\frac{1}{(2 \pi)^2}\int d^{3}p\, \epsilon_{ijk}\mathcal{\tilde{H}}^{ijk},
\end{eqnarray}
which is a topological $\mathbb{Z}$ number and, in fermion systems, it coincides with the topological invariant of chiral topological insulators.
We can finally combine Eq. (\ref{string-equations}) with Eq. (\ref{potential}) to obtain
\begin{eqnarray}
	\dot{r}^{i}=\tilde{\mathcal{H}}^{ijk} E_{jk}+...,
\end{eqnarray}
where the dots denote the term $\partial H/ \partial p_i$ that we have omitted, as it does not contribute to the topological response of the system. Due to the definition of the current $J^{i}=1/(2\pi)^3\int d^{3}p\, \dot{r}^{i}$, which is the three-dimensional version of Eq. (\ref{current2}), we integrate in momentum space both terms of above equation and find
\begin{eqnarray}
	J^{i}= \frac{\kappa}{2 \pi}\epsilon^{ijk}E_{jk},
\end{eqnarray}
which exactly matches the vector current derived in Eq. (\ref{stringQHE}) from the BF theory. This represents the central result of this section and shows how the response of a topological system that supports extended objects can also be derived at the semiclassical level by extending the phase-space formalism to quasi-strings.
Similar results can be naturally achieved for higher-dimensional quasi-branes.

\section* {\bf Conclusions and outlook}

\noindent In this work, we have introduced and discussed a semiclassical phase-space approach for quasi-strings in the context of topological phases of matter. We have presented a generalized Hamiltonian formalism for extended objects that has allowed us to derive a quantized topological current, consistent with a complementary derivation from a topological BF theory. Although we have primarily focused on one-dimensional extended objects embedded in three spatial dimensions, such as vortex strings in superfluid phases, our formalism can be naturally generalized to higher-dimensional extended objects, like quasi-membranes, which are relevant, for instance, in the quantum Hall effect in higher dimensions. Importantly, we have shown that tensor Berry connections play a crucial role in deriving the topological current associated with the endpoints of open quasi-strings, which reside on the boundary of the system and support an quantized Hall response. One natural future direction will be shifting from the semiclassical to the fully quantum regime within the context of deformation quantization for extended objects, showing, for instance, the existence of novel quantum effects in the presence of strong magnetic fields. We believe that our approach will shed light on the emergence of collective modes and non-commutative geometry in interacting topological matter by generalizing some of the results obtained in the fractional quantum Hall effect for quasi-particles in two spatial dimensions and in other higher-dimensional topological phases \cite{Neupert,Hasebe,Chu2,Ryu}. Finally, we foresee a possible relationship between quantum volume-preserving diffeomorphisms of (interacting) extended objects in higher-dimensional topological matter and some results related to non-commutative geometry on D-branes \cite{Chu1999,Jabbari,Schomerus,Bergshoeff,Matsuo,Pioline,Szabo}.


\bibliography{references}

\end{document}